\newif\ifanonymized
\anonymizedtrue     
\anonymizedfalse    

\newif\ifsubmit
\submitfalse	
\submittrue	

\documentclass[sigconf,pbalance]{acmart}

\usepackage{fancyhdr}

\usepackage{url}

\usepackage{color}

\usepackage{xspace}

\usepackage{graphicx}
\DeclareGraphicsExtensions{.pdf,.png,.eps,.ps,.jpg}

\usepackage{multirow}

\newcommand{\bh}{-\allowbreak}

\ifsubmit
  \newcommand{\hey}[1]{\relax}
  \newcommand{\heyvarun}[1]{\relax}
  \newcommand{\bibtex}[1]{\relax}
  \newcommand{\note}[1]{\relax}
\else
  \newcommand{\hey}[1]{\textcolor{magenta}{[{#1}]}}
  \newcommand{\heyvarun}[1]{\textcolor{blue}{[Varun: {#1}]}}
  \newcommand{\note}[1]{\par\textcolor{magenta}{Note: {#1}}\par}
  \newcommand{\bibtex}[1]{\textcolor{red}{@bibtex}\{#1\}} 
\fi

\newcommand{\hide}[1]{\relax}

\newcommand{\figlabel}[1]{\label{fig:#1}}
\newcommand{\figref}[1]{Figure~\ref{fig:#1}}

\newcommand{\addfigure}[4]{
\begin{figure}[tbp]
\centerline{\resizebox{#2\linewidth}{!}{\includegraphics{figs/#1}}}
\caption{\figlabel{#1}#3}
\Description{#4}
\end{figure}
}

\newcommand{\tablabel}[1]{\label{tab:#1}}
\newcommand{\tabref}[1]{Table~\ref{tab:#1}}

\newcommand{\Fpt}{$\mathbf{F}_{\mathrm{pt}}$}
\newcommand{\Fbase}{$\mathbf{F}_{\mathrm{base}}$}
\newcommand{\Fglobal}{$\mathbf{F}_{\mathrm{global}}$}

\newcommand{\seclabel}[1]{\label{sec:#1}}
\newcommand{\secref}[1]{Section~\ref{sec:#1}}

\copyrightyear{2025}
\acmYear{2025}
\setcopyright{cc}
\setcctype{by}
\acmConference[UbiComp Companion '25]{Companion of the 2025 ACM International Joint Conference on Pervasive and Ubiquitous Computing}{October 12--16, 2025}{Espoo, Finland}
\acmBooktitle{Companion of the 2025 ACM International Joint Conference on Pervasive and Ubiquitous Computing (UbiComp Companion '25), October 12--16, 2025, Espoo, Finland}\acmDOI{10.1145/3714394.3756241}
\acmISBN{979-8-4007-1477-1/2025/10}

\begin{document}

\title{Beyond Motion Artifacts: Optimizing PPG Preprocessing for Accurate Pulse Rate Variability Estimation}
\date{}

\ifanonymized
    \author{Anonymized for blind submission}
\else
    \author{Yuna Watanabe}
    \orcid{1234-5678-9012}
    \affiliation{
     \institution{Northeastern University}
     \city{Boston}
     \state{MA}
     \country{USA}}
    \email{watanabe.y@northeastern.edu}
    \author{Natasha Yamane}
    \orcid{0000-0002-3532-1563}
     \affiliation{
      \institution{Northeastern University}
      \city{Boston}
      \state{MA}
      \country{USA}}
    \email{yamane.n@northeastern.edu}
    \author{Aarti Sathyanarayana}
    \orcid{0000-0003-1351-2315}
     \affiliation{
      \institution{Northeastern University}
      \city{Boston}
      \state{MA}
      \country{USA}}
    \email{a.sathyanarayana@northeastern.edu}
    \author{Varun Mishra}
    \orcid{0000-0003-3891-5460}
    \authornote{These authors jointly supervised this work and share senior authorship.}
     \affiliation{
      \institution{Northeastern University}
      \city{Boston}
      \state{MA}
      \country{USA}}
    \email{v.mishra@northeastern.edu}
    \author{Matthew S. Goodwin}
    \orcid{0000-0002-4237-601X}
    \authornotemark[1]
     \affiliation{
      \institution{Northeastern University}
      \city{Boston}
      \state{MA}
      \country{USA}}
    \email{m.goodwin@northeastern.edu}
\fi



\begin{abstract}
Wearable physiological monitors are ubiquitous, and photoplethysmography (PPG) is the standard low\bh cost sensor for measuring cardiac activity. Metrics such as inter\bh beat interval (IBI) and pulse\bh rate variability (PRV)---core markers of stress, anxiety, and other mental\bh health outcomes---are routinely extracted from PPG, yet preprocessing remains non\bh standardized. Prior work has focused on removing motion artifacts; however, our preliminary analysis reveals sizeable beat\bh detection errors even in low\bh motion data, implying artifact removal alone may not guarantee accurate IBI and PRV estimation. We therefore investigate how band\bh pass cutoff frequencies affect beat\bh detection accuracy and whether optimal settings depend on specific persons and tasks observed. We demonstrate that a fixed filter produces substantial errors, whereas the best cutoffs differ markedly across individuals and contexts. Further, tuning cutoffs per person and task raised beat\bh location accuracy by up to 7.15\% and reduced IBI and PRV errors by as much as 35~ms and 145~ms, respectively, relative to the fixed filter. These findings expose a long\bh overlooked limitation of fixed band\bh pass filters and highlight the potential of adaptive, signal\bh specific preprocessing to improve the accuracy and validity of PPG\bh based mental\bh health measures.
\end{abstract}

\ccsdesc[500]{Human-centered computing~Ubiquitous and mobile computing systems and tools}

\keywords{Physiological signal processing; Wearable sensing; Photoplethysmography; Mental health}

\maketitle  


\ifsubmit
    \relax
\else
    \par\noindent \textcolor{red}{\textbf{DRAFT}: \today\ -- \currenttime}
    \pagestyle{fancy}
    \lhead{DRAFT in preparation}
    \rhead{version: \today\ -- \currenttime}
    \chead{}
    \lfoot{}
    \rfoot{}
\fi
\section{Introduction} 
\seclabel{introduction}
Recent advances in wearable peripheral sensors allow researchers to continuously and non\bh invasively monitor physiological states. Photoplethysmography (PPG) has emerged as a popular method for monitoring cardiac activity because of its affordability, portability, and ease of integration in consumer devices. In the context of mental and behavioral health, PPG signals have been leveraged for myriad purposes, including, but not limited to, stress detection \cite{tazarv_personalized_2021, park_study_2018, arsalan_human_2021, zangroniz_estimation_2018, toshnazarov_sosw_2024, mishra_investigating_2018, mishra_evaluating_2020}, depression monitoring \cite{unursaikhan_development_2021}, aggression onset prediction \cite{imbiriba_wearable_2023}, and detecting craving for substances \cite{chintha_wearable_2018, holtyn_towards_2019}. These analyses often rely on inter-beat interval (IBI) and pulse rate variability (PRV) that reflect autonomic nervous system regulatory activity. Accurate IBI and PRV estimates are therefore critical---any bias at this stage propagates to downstream models, undermining prediction accuracy and validity in mental\bh health applications.

Despite its widespread use, PPG signal preprocessing remains non-standardized \cite{dunn_building_2025}. Although most pipelines begin with a band\bh pass filter to remove low- and high\bh frequency noise unrelated to cardiac activity, the choice of cutoff frequencies varies widely across studies. Prior work has attempted to identify optimal cutoffs \cite{wolling_optimal_2021,liang_optimal_2018,mejia-mejia_effects_2023,mejia-mejia_effect_2021}, with some proposing 0.5 to 15.0~Hz as a suitable, universal passband~\cite{wolling_optimal_2021}; however, researchers have not reached a consensus. Further, optimal filtering parameters depend on individual factors (e.g., age, baseline heart rate), sensor placement, and the nature of the task being recorded \cite{cassani_optimal_2020,wolling_optimal_2021}. Given this complexity, researchers are often left to select filtering configurations that best fit their datasets, which can be arbitrary and inconsistent. Some work has investigated more adaptive denoising techniques instead of, or in addition to, applying a fixed filter, including least\bh mean\bh squares (LMS) filtering \cite{ram_novel_2012, peng_motion_2014}, wavelet denoising \cite{joseph_photoplethysmogram_2014}, and deep learning\bh based approaches \cite{ kwon_preeminently_2022, xu_photoplethysmography_2020, azar_deep_2021}. However, these methods are typically designed only to reduce motion artifacts and may not generalize well to other sources of PPG signal distortion \cite{kleckner_framework_2021}. These issues raise intertwined questions: Is adaptively denoising motion artifacts enough to yield accurate IBI and PRV estimates? If not, how much does parameterization in band\bh pass filtering matter, and can any single fixed filter suffice?


This paper aims to evaluate whether a one\bh size\bh fits\bh all band\bh pass filter yield accurate IBI and PRV estimates, and, if not, quantify how much person-, activity-, and mental\bh state–specific optimization of cutoff frequencies improves accuracy. In pursuing these goals, our work generates three key insights:
\begin{itemize}
    \item Beyond physical activity, PPG signals contain inherent noise influenced by the individual, their specific activities, and mental states.
    \item Applying a fixed band-pass filter can introduce substantial error in beat detection and resulting IBI and PRV estimates.
    \item Choosing activity- and state\bh specific cutoff frequencies for each individual significantly reduces beat-detection errors and improves IBI and PRV estimation.
\end{itemize}
Our findings suggest that preprocessing pipelines should be tailored to each individual and recording context to obtain accurate IBI and PRV estimates, beyond denoising motion artifacts. Doing so can enhance the validity of physiological metrics used in mental and behavioral health monitoring and lay the groundwork for more personalized and accurate assessment tools.

\section{Background \& Related Work} 
\seclabel{background}


\subsection{Preprocessing and filtering strategies}
\seclabel{background-preprocessing-methods}
Researchers routinely leverage PPG signals to identify various affective states, including stress and depression, commonly deriving metrics like IBI (or pulse rate) and PRV. Because filtering strategies depend on the specific metric of interest, we focus on preprocessing methods to extract IBI or PRV from PPG signals.

Although considerable variability in filtering configurations exists, band\bh pass filtering is the most commonly used preprocessing technique~\cite{tazarv_personalized_2021, deviaene_sleep_2018, park_study_2018, unursaikhan_development_2021,  zangroniz_estimation_2018}. Low\bh cut frequencies used in prior work range from 0~Hz (i.e., no low\bh pass filtering) to 0.9~Hz, while high-cut frequencies span 1.6 to 35~Hz.

In addition to band\bh pass filtering, some researchers apply moving averages \cite{tazarv_personalized_2021} and wavelets \cite{unursaikhan_development_2021}, while others opt for Savitzky–Golay filters \cite{arsalan_human_2021}. While a few studies briefly explain their rationale for filtering (e.g., to remove non\bh cardiac signals) \cite{tazarv_personalized_2021, park_study_2018, unursaikhan_development_2021, zangroniz_estimation_2018}, most do not justify their choice of filtering methods or cutoff frequencies, suggesting a lack of consensus on optimal preprocessing practices. This observation reinforces our conjecture that filtering parameters are often chosen heuristically rather than following standardized guidelines \cite{kleckner_framework_2021}.

Given the lack of standardized preprocessing practices, several researchers have attempted to identify optimal filtering strategies. For example, Liang et al. \cite{liang_optimal_2018} compared seven filtering algorithms with various filter orders applied to short\bh duration PPG signals and found that a 4th\bh order Chebyshev Type\bh II filter yielded the best improvement in signal quality indices. Other studies have focused on determining the optimal cutoff frequencies for band\bh pass filters (\tabref{background-best-preprocessing}). Despite these efforts, different studies report wide variation in recommended frequency ranges, with no clear agreement.

\begin{table*}[tbp]
\caption{\tablabel{background-best-preprocessing}Summary of recommended low\bh cut and high\bh cut frequencies in prior work.}
\centerline{\resizebox{1.0\linewidth}{!}{\begin{tabular}{ l l l l l }
\toprule
 Publication & Low-cut & High-cut & Type of filter & Evaluation \\
 \midrule
 Cassani et al. \cite{cassani_optimal_2020} & 
 0.6~Hz (adult); 1.0~Hz (child) & 
 3.3~Hz (adult); 2.7~Hz (child) & 
 Bandpass filter & 
 Signal-to-noise ratio\\
 
 Karolcik and Georgious \cite{karolcik_optimal_2024} & 0.35 Hz & 10 Hz & 4th-order Chebyshev Type-II & Signal quality indices \\
 Mejía-Mejía et al. \cite{mejia-mejia_effect_2021} & Around 0 Hz & - & 4th-order Butterworth filter & PRV accuracy \\
 Wolling et al. \cite{wolling_optimal_2021} & 0.5 Hz & 15 Hz &  & Peak displacement\\
 Mejía-Mejía and Kyriacou \cite{mejia-mejia_effects_2023} & Around 0 Hz & 20 Hz & FIR and IIR filters & PRV accuracy\\  
 \bottomrule
\end{tabular}}}
\end{table*}

\subsection{Adaptive denoising techniques}
\seclabel{background-denoising}

In addition to using fixed band\bh pass filters, researchers have explored various adaptive techniques to denoise PPG signals. Classic approaches include Fourier series analysis \cite{reddy_use_2009} and wavelet denoising \cite{joseph_photoplethysmogram_2014}, which decompose motion\bh corrupted signals into frequency components, remove noise, and reconstruct cleaner PPG waveforms. Sparse signal reconstruction has also been used to isolate relevant signal components while discarding noise \cite{zhang_troika_2015}.

Another widely studied method involves LMS\bh based adaptive filters \cite{ram_novel_2012, peng_motion_2014} that optimize filter coefficients to minimize error between the estimated clean and observed signal. While some LMS\bh based approaches work without additional sensors by reconstructing noise from PPG itself \cite{ram_novel_2012}, most rely on reference noise sources such as accelerometer data. Independent component analysis provides an alternative that separates underlying signal sources without requiring additional sensors \cite{peng_motion_2014, kim_motion_2006, krishnan_two-stage_2010}.

Recent work has applied machine learning\bh based denoising methods, including generative models \cite{kwon_preeminently_2022}, time\bh delay neural networks \cite{xu_photoplethysmography_2020}, and autoencoder architectures \cite{azar_deep_2021} that aim to learn signal\bh cleaning functions directly from data.

Although these methods show promise in mitigating motion artifacts, they primarily focus on motion\bh related noise. In our preliminary analysis (\secref{non-motion-artifacts}), we explore whether motion artifacts are a single noise source that impacts beat\bh detection accuracy.

\section{Methods} 
\seclabel{methods}
To evaluate the impacts of filtering parameters on beat\bh detection accuracy, we apply different combinations of cutoffs and compare their performance. We use two datasets to account for variability between different experimental configurations. 

\subsection{Datasets}
\seclabel{dataset}
We used the Wearable Stress and Affect Detection (WESAD) and Stress\bh free datasets, which include ECG and PPG signals collected simultaneously. This allows us to use ECG\bh derived beats as ground truth when evaluating the validity of PPG\bh derived beats.

\textbf{WESAD dataset:} The publicly available WESAD dataset was developed for stress and affect detection \cite{schmidt_introducing_2018}. It contains ECGs collected with the RespiBAN chest sensor and PPGs collected with the Empatica E4 wristband. Fifteen participants performed four tasks to induce different affective states: baseline (20~minutes), amusement (5~minutes), stress (10~minutes), and meditation (7~minutes $\times$ 2). Although the data includes transient periods during which participants completed self\bh reports, we only used data from the four tasks. 

\textbf{Stress-free dataset:} The Stress\bh free dataset was collected to advance stress detection \cite{amin_extending_2025}. The dataset contains beats from ECGs collected with the Polar H10 and PPGs collected with the Empatica E4 wristband. Thirty\bh five participants underwent three stressful tasks after a 10\bh minute baseline rest period: mental arithmetic, startle response, and a cold pressor task, each lasting 4 minutes. Participants were given a 5\bh minute rest period between each task, wherein they remained seated without engaging in any activity; however, we use only the baseline rest data collected at the beginning and the three stress\bh inducing tasks in our analyses. 

\subsection{Initial data preprocessing}
\seclabel{preprocessing}
We used the heartview package, a Python package for preprocessing signals \cite{yamane_heartview_2024}. We preprocessed WESAD ECG signals using an elliptic band\bh pass filter (1--15~Hz) and Manikandan and Soman's \cite{manikandan_novel_2012} beat\bh detection algorithm. Following Liang et al. \cite{liang_optimal_2018}, we preprocessed PPG data with a 4th\bh order Chebyshev Type\bh II filter. We filtered PPGs using 525 parameter combinations: low\bh cut frequencies ($f_{C,\text{low}}$; 0.4–1.7~Hz) and high\bh cut frequencies ($f_{C,\text{high}}$; 1.2–5.0~Hz), both in 0.1-Hz increments, excluding combinations where $f_{C,\text{low}}$ was larger than $f_{C,\text{high}}$. We used HeartPy's \cite{van_gent_heartpy_2019} algorithm to detect beats in PPGs, defining each beat location as the middle\bh amplitude point between foot and apex to reduce IBI and PRV estimation errors \cite{peralta_optimal_2019}.

\subsection{Metrics for evaluation}
\seclabel{metrics}
 We evaluate three accuracy metrics on PPG signals filtered with our 525 unique filter combinations: (1) beat location accuracy (compared to ECG peaks); (2) IBI accuracy (compared to ECG\bh derived R-R intervals); and (3) PRV accuracy (compared to ECG\bh derived heart rate variability [HRV]). We then visualize how accuracy metrics vary with filtering parameters and examine differences in optimal settings across persons and tasks. 


\textbf{Beat location accuracy:} Beat location accuracy measures how closely PPG beat locations match those of ECG beats. Following Charlton et al. \cite{charlton_establishing_2022}, we label a PPG beat ``correct" when it occurs within 150~ms of the nearest ECG beat. Owing to potential misalignments in sampling rate, device clocks, and waveform shape, Charlton et al. searched offsets between PPGs and ECGs up to $\pm 10$ s. Visual inspection showed our recordings drifted far less, so we limited our lag search to $\pm 2$ s in 20~ms steps and retained offsets that maximized matches. We then evaluate improvements in sensitivity ($Se$), positive predictive value ($PPV$), and F1 score (F1). Each score was calculated using the following formulas:
\[
    Se=\frac{n_\mathrm{correct}}{n_\mathrm{ECG}} \times 100,
    \quad
    PPV=\frac{n_\mathrm{correct}}{n_\mathrm{PPG}} \times 100,
    \quad
    F1=\frac{2 \times PPV \times Se}{PPV + Se}
\]
where $n_\mathrm{correct}$ is the number of correct PPG peaks, $n_\mathrm{ECG}$ is the total number of ECG\bh derived peaks, and $n_\mathrm{PPG}$ is the total number of PPG\bh derived peaks. A high F1 score indicates that the selected filtering parameters yield optimal accuracy in beat detection.


\textbf{IBI accuracy:} IBI accuracy measures how closely PPG\bh derived IBIs match ground\bh truth ECG\bh derived R-R intervals. We segmented the signals into 60-second windows and removed artifactual beats using the FLIRT package to avoid skewing mean IBI estimates \cite{foll_flirt_2021}. For segments with more than 10 valid beats, we computed the average IBI and the difference between R-R intervals and IBIs. We then computed the mean absolute error (MAE) of intervals per participant and task. Lower MAE indicates higher IBI accuracy, reflecting better beat detection.


\textbf{PRV accuracy:} As with IBI, we assessed PRV accuracy by comparing PPG\bh derived PRV with ECG\bh derived HRV. We used RMSSD (root mean square of successive IBI differences) as the metric for variability. Following the same procedure for IBI accuracy, we divided signals into 60-second segments, removed artifactual beats, and calculated RMSSD. We then calculated the difference between ECG- and PPG\bh derived values and computed the MAE per participant and task. Lower MAE indicates higher PRV accuracy.


To further evaluate the potential impact of signal\bh by\bh signal adaptive preprocessing, we identify optimal filter settings at two levels: (1) across all tasks and participants (\Fglobal) and (2) per participant within each task (\Fpt). We ran multi\bh objective optimization at each level using the NSGA-II algorithm in the `pymoo' package \cite{blank_pymoo_2020} with average F1 score, MAE IBI, and MAE RMSSD as the multiple objectives. NSGA‐II may return multiple solutions for which none is strictly better in all three objectives. We chose the single best filter by performing min\bh max normalization for each objective and minimizing the combined metric $-\mathrm{F1_{\mathrm{norm}}} \;+\;\mathrm{MAE}_{\mathrm{IBI, norm}}\;+\;\mathrm{MAE}_{\mathrm{RMSSD, norm}}$. As a baseline, we also apply a fixed non\bh optimized band\bh pass filter (0.5--4.0Hz; \Fbase), as these values align with commonly recommended ranges (\secref{background-preprocessing-methods}). We then compare the resulting average performance (F1, MAE IBI, and MAE RMSSD) across the three filter settings (\Fbase, \Fglobal, and \Fpt). Finally, we compare the distribution of mean IBI and RMSSD against the ECG reference across \Fbase, \Fglobal, and \Fpt{} using a repeated\bh measures ANOVA ($\alpha=.95$) followed by Bonferroni\bh corrected pairwise t-tests.

\section{Findings and Evaluation}
\seclabel{findings-evaluation}
\subsection{The non-``motion'' artifacts}
\seclabel{non-motion-artifacts}

We first ran a preliminary correlation analysis to test whether motion artifacts fully explain beat\bh detection errors. We preprocessed PPG signals with a fixed 0.5--4~Hz band\bh pass filter, segmented the signal into 60-second windows, and computed beat\bh detection accuracy metrics (\secref{metrics}). We calculated Monitor Independent Movement Summary unit (MIMS-unit)---a metric capturing human motion from accelerometer signals \cite{john_open-source_2019}---and used its Area\bh Under\bh the\bh Curve ($\mathrm{AUC}_\mathrm{MIMS}$) to quantify motion artifact level.

We show the correlation between motion level ($\mathrm{AUC}_\mathrm{MIMS}$) and beat-detection accuracy for four representative low\bh motion tasks in \figref{mims-vs-beat-accuracy}. While some other tasks show moderate\bh to\bh strong correlations, low\bh motion tasks reveal many 60-second segments with small $\mathrm{AUC}_\mathrm{MIMS}$ yet poor accuracy (low F1; high MAE IBI and RMSSD). This finding suggests that beat\bh detection errors can persist even when motion artifacts are minimal, motivating our search for a more flexible filtering strategy.

\addfigure{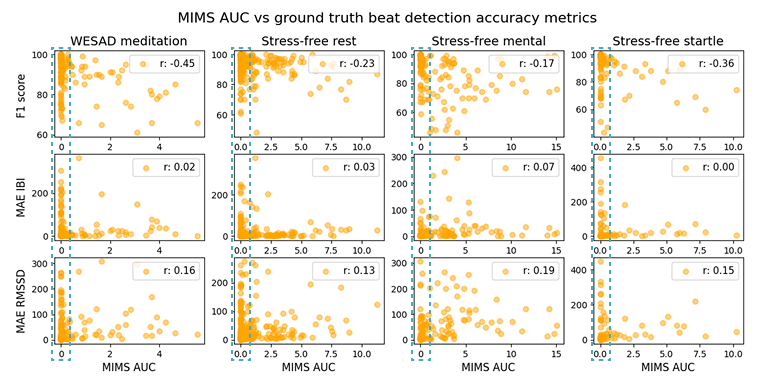}{1.0}{Correlations between motion artifacts ($\mathrm{AUC}_\mathrm{MIMS}$) and beat\bh detection accuracy (F1 score, MAE IBI \& RMSSD).}{Pearson's correlations between $\mathrm{AUC}_\mathrm{MIMS}$ and F1 score, MAE IBI, and MAE RMSSD are small to moderate, ranging from 0.00 to 0.45.}

\subsection{Evaluating different parameters}
\seclabel{eval-different-params}

\addfigure{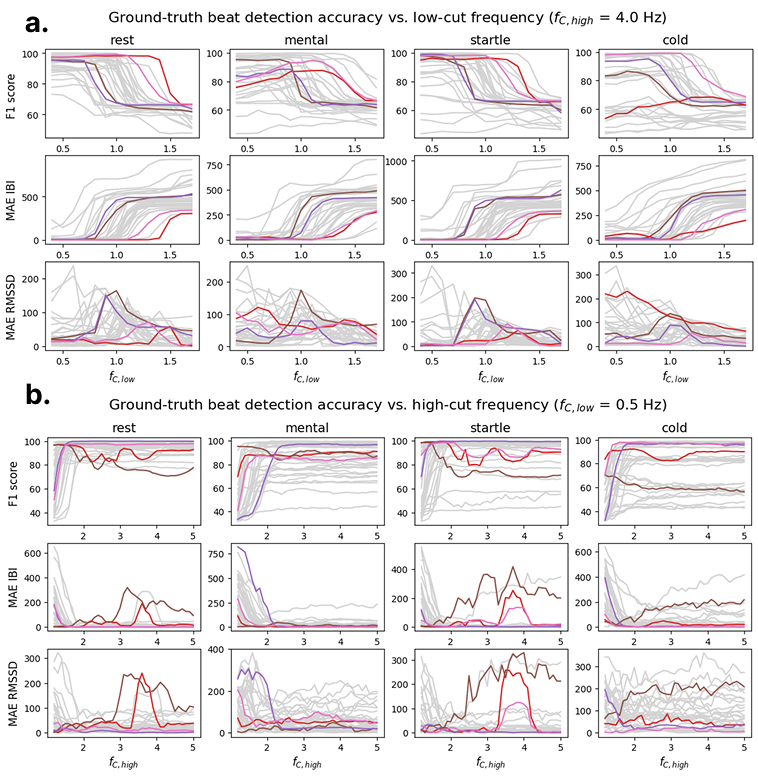}{1.0}{F1 score and MAE IBI and RMSSD changes across varying (a) low\bh cut and (b) high\bh cut frequencies in the bandpass filter for the Stress\bh free dataset. Each column represents a task, and each line represents a single participant's metric.}{}

We present F1 scores and MAE IBI and RMSSD across signal segments at varying $f_{C,\text{low}}$ values for the Stress\bh free dataset in \figref{filter-vs-beat-accuracy}a. The choice of $f_{C,\text{low}}$ impacted every metric in both datasets. F1 scores drop, and MAE IBI and RMSSD increase as $f_{C,\text{low}}$ varied between 0.5 and 1.5~Hz (i.e., the typical human heart rate range); however, these frequencies vary across individuals. For example, with $f_{C,\text{high}}$ of 4.0~Hz, participants represented by pink and red lines exhibited peak F1 scores around a $f_{C,\text{low}}$ of 1.2~Hz, while participants represented by brown and purple lines peaked around 0.8~Hz. Although some participants showed stable F1 scores and MAE at lower $f_{C,\text{low}}$ (0.5--1.5~Hz), others were sensitive to slight $f_{C,\text{low}}$ changes (e.g., pink and red lines in the mental stress panels).

The choice of $f_{C,\text{high}}$ also affected beat-detection accuracy (\figref{filter-vs-beat-accuracy}b), though less than $f_{C,\text{low}}$. Selecting a lower $f_{C,\text{high}}$ (below 2~Hz) generally lowered F1 scores and raised MAE IBI and RMSSD; however, higher $f_{C,\text{high}}$ impaired performance for some participants and tasks (e.g., red and brown lines in the rest and startle response panels). The optimal $f_{C,\text{high}}$ also varied across participants and tasks. While \figref{filter-vs-beat-accuracy}a-b only show cases where $f_{C,\text{high}}=4.0 \text{Hz}$ and $f_{C,\text{low}}=0.5 \text{Hz}$, respectively, varying these parameters also affected the graph waveform, indicating a complex relationship between the them. These results suggest that an arbitrary choice may significantly impact beat\bh detection accuracy for some participants and tasks, indicating the need for individual- and task\bh specific filter parameterization. The WESAD dataset produced similar trends.

\addfigure{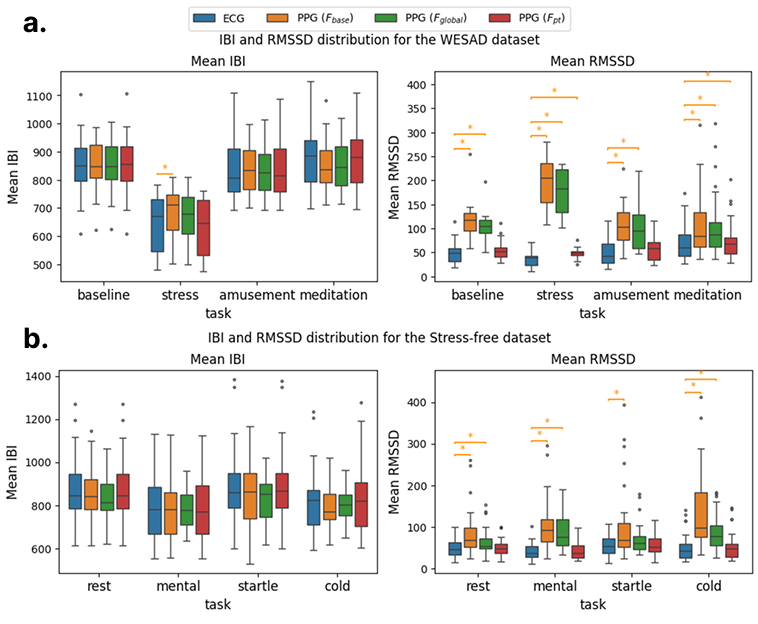}{1.0}{Distribution of mean IBI and RMSSD derived from ECGs and PPGs preprocessed with three types of filters for (a) the WESAD and (b) the Stress\bh free datasets.  Brackets with asterisks indicate significant pairwise differences.}{Distribution of mean RMSSD between ECG and PPG (\Fbase) and PPG (\Fglobal) are significantly different, while the distribution of PPG (\Fpt) is much closer to that of ECG. The difference is especially significant in the WESAD stress task and the Stress\bh free cold task. \Fglobal{} is slightly closer to ECG than \Fbase, however, it still deviates significantly.}

\begin{table*}[tbp]
\caption{\tablabel{optimization-results}Mean F1 score and MAE in IBI and RMSSD with different types of cutoff frequency optimization. Optimization per task and person produced the smallest errors in IBI and RMSSD estimations.}
\centerline{\resizebox{1.0\linewidth}{!}{\begin{tabular}{ l l | r r r | r r r | r r r }
\toprule
 \multirow{2}{5em}{Dataset} & \multirow{2}{5em}{Task} 
   & \multicolumn{3}{c}{Fixed ($\mathbf{F}_{\mathrm{base}}$)} 
   & \multicolumn{3}{c}{Optimized across task \& person ($\mathbf{F}_{\mathrm{global}}$)} 
   & \multicolumn{3}{c}{Optimized per task \& person ($\mathbf{F}_{\mathrm{pt}}$)} \\
 & 
   & Mean F1 & MAE IBI & MAE RMSSD 
   & Mean F1 & MAE IBI & MAE RMSSD 
   & Mean F1 & MAE IBI & MAE RMSSD \\ 
\midrule
\multirow{4}{5em}{WESAD} 
 & baseline   & 69.95 & 17.94 & 68.10 & 71.12 & 15.47 & 55.31 & \textbf{71.72} & \textbf{4.470} & \textbf{6.380} \\ 
 & stress     & 60.96 & 43.36 & 158.7 & 62.64 & 42.49 & 139.98 & \textbf{68.11} & \textbf{15.97} & \textbf{13.34} \\ 
 & amusement  & 84.54 & 24.38 & 59.60 & 85.37 & 23.00 & 46.64 & \textbf{87.05} & \textbf{3.655} & \textbf{8.729} \\ 
 & meditation & 90.00 & 29.30 & 40.58 & 90.36 & 28.32 & 34.03 & \textbf{92.01} & \textbf{4.830} & \textbf{6.409} \\ 
\hline
\multirow{4}{5em}{Stress-free} 
 & cold      & 77.21 & 46.91 & 85.88 & 77.92 & 63.87 & 39.41 & \textbf{83.02} & \textbf{11.95} & \textbf{8.188} \\ 
 & mental    & 83.23 & 20.30 & 61.80 & 82.17 & 41.77 & 44.78 & \textbf{89.03} & \textbf{3.497} & \textbf{5.128} \\ 
 & rest      & 93.31 & 16.20 & 36.31 & 94.13 & 37.38 & 18.18 & \textbf{97.25} & \textbf{1.156} & \textbf{2.011} \\ 
 & startle   & 89.65 & 28.88 & 44.84 & 89.62 & 51.21 & 20.25 & \textbf{93.94} & \textbf{2.835} & \textbf{3.031} \\ 
\bottomrule
\end{tabular}}}
\end{table*}

\tabref{optimization-results} reports \emph{mean} performance across participants; in both datasets, \Fpt{} achieved the highest F1 scores and the lowest IBI and RMSSD errors. Relative to \Fbase{}, the largest mean gains were a \textbf{7.15\%} in F1 (WESAD stress), a \textbf{35~ms} reduction in MAE IBI (Stress\bh Free cold), and a \textbf{145~ms} reduction in MAE RMSSD (WESAD stress). Comparing within \emph{individuals} reveals still larger benefits:  
F1 improved by up to \textbf{28.8\%} (median 2.50\%, IQR 0.35--6.73\%), while MAE IBI and MAE RMSSD decreased by as much as \textbf{328\;ms} (median 2.96\;ms, IQR 0.15--14.9\;ms) and \textbf{329\;ms} (median 26.5\;ms, IQR 8.65--73.1\;ms), respectively. We also show the mean IBI and RMSSD distribution in \figref{ibi-rmssd-dist}. Repeated\bh measures ANOVA revealed a significant mean IBI difference between ECGs and PPGs only in WESAD stress; other tasks showed none, suggesting that using generic or fixed filtering thresholds may be sufficient for estimating aggregated/high\bh level metrics about heart rate or IBI. In contrast, mean RMSSD differed across all tasks, indicating that fixed thresholds introduce substantial errors when measuring variability---a key indicator of autonomic nervous system activity and variable of interest for a variety of mental and behavioral health outcomes. Specifically, mean RMSSD derived from the PPGs preprocessed with \Fbase{} and \Fglobal{} showed significantly different distributions compared to ECGs in almost all tasks, with medium\bh to\bh large effect sizes (Cohen's d=0.420--4.04). PPG signals preprocessed with \Fpt{} also showed a significant RMSSD difference in the WESAD stress task, \emph{p} = .0007, and meditation task, \emph{p} = .0074. However, effect sizes were markedly smaller than for \Fbase{} and \Fglobal: stress task---Cohen’s $d =$ 4.04 (\Fbase), 3.90 (\Fglobal), and 0.96 (\Fpt); meditation task---$d =$ 0.71 (\Fbase), 0.62 (\Fglobal), and 0.11 (\Fpt). These results demonstrate that fixed cutoffs lead to substantial errors, and signal\bh by\bh signal filtering optimization can significantly enhance the accuracy and validity of beat detection and beat variability estimation.

\section{Discussion}
\seclabel{discussion}


\subsection{Impact of cutoff frequencies on RMSSD}
\seclabel{discussion-rmssd}

In our analyses (\secref{eval-different-params}), although F1 scores and MAE IBI exhibited inverse trends, MAE RMSSD followed a distinct pattern: it increased at specific low\bh cut frequencies and then decreased again. This behavior may be explained by the nature of the PPG waveform, which often exhibits two prominent peaks per cardiac cycle---one corresponding to the systolic wave and the other to the diastolic wave. As the low\bh cut frequency increases, low\bh frequency components are attenuated, and higher\bh frequency components are emphasized. Consequently, the beat\bh detection algorithm may erroneously identify both systolic and diastolic peaks as separate beats, doubling the beat count. Interestingly, this consistent misidentification introduces a form of regularity in the detected IBIs, resulting in RMSSD values that appear more stable and, in some cases, closer to those from correctly identified beats. This contrasts with intermediate low\bh cut frequencies, where beat detection is less consistent, leading to greater IBI variability and larger RMSSD errors.

\subsection{Factors influencing optimal cutoffs}
\seclabel{discussion-factors}

Heart rate is one of the most apparent factors affecting optimal cutoff frequencies, which vary with age, lung capacity, and activity level. If heart rate alone could determine optimal filtering parameters, applying frequency cutoffs well outside the plausible heart rate range would preserve signal quality. However, our results demonstrate that this is not always the case---some PPG signals remain highly sensitive to cutoff frequency settings, and broader passbands do not always yield accurate beat detection.

Additional factors may include stress and its physiological effects. Prior studies have shown that stress can alter PPG waveform morphology \cite{celka_influence_2019, rinkevicius_photoplethysmogram_2019}. Mental stress induces vasodilation in the forearm \cite{lindqvist_sustained_1996}, affecting PPG amplitude. Respiration also modulates PPG amplitude, wherein inhalation reduces it, and exhalation amplifies it. These morphological changes may influence the effectiveness of different filters. Indeed, we observed that PPG signals collected during the mental arithmetic and cold pressor tasks in the Stress\bh free dataset and the stress task in the WESAD dataset were more sensitive to low\bh cut frequencies than those collected during rest. This suggests that physiological changes induced by the person's activities and mental states may impact optimal configurations.

A limitation of the present work is that we employed a single beat\bh detection algorithm, yet choice of algorithm may influence optimal cutoff frequencies. Some detectors are likely more tolerant of small shifts in filter settings, whereas others require carefully tuned parameters. We also only examined data from Empatica E4 wristbands collected in the lab. We believe that similar patterns can be observed in free\bh living data as well, once motion artifacts are removed. However, different devices may apply different preprocessing to raw data, and different on\bh body placement locations may exhibit distinct waveform patterns. Future work should investigate how different algorithms, devices, and sensor placements interact with preprocessing choices across in-lab and in-the-wild data.

\subsection{The need for adaptive filtering}
\seclabel{discussion-future-work}

Overall, our findings suggest that cutoff frequencies significantly impact beat\bh detection accuracy when motion artifacts are negligible, and the optimal configuration may vary across individuals, activities, and mental states. This dependence has been largely overlooked: applying a conventional fixed filter can distort IBI and RMSSD in a context- and person\bh specific manner. Echoing electrodermal work showing that adaptive thresholding enhances sensitivity \cite{kleckner_adaptive_2024}, these observations call for adaptive preprocessing for accurate pulse rate variability estimation. During model development, researchers may be able to manually verify that a fixed filter behaves sensibly; however, once deployed in the field, they cannot inspect every new signal. If the filter systematically underestimates or overestimates PRV under stress for a user, for example, the model may misclassify that state. Tuning the filter to each signal---reflecting the user’s physiology and situation---avoids systematic errors that would otherwise propagate through the pipeline, degrading the accuracy and validity of downstream mental- and behavioral\bh health models.

In this study, we tuned filter parameters using reference ECGs. However, such ground-truth signals are often unavailable in real-world applications. Future work should develop task- and person\bh aware optimization methods that operate without ECG references.

\section{Conclusion}
\seclabel{conclusion}

We investigate how non‑motion noise in PPG signals degrades beat\bh detection accuracy and how filter settings affect that error. We demonstrate that PPG contains noise beyond physical movement, so motion\bh artifact removal alone may not ensure accurate and valid IBI and PRV estimates. We also demonstrate that using a single band\bh pass filter for all individuals, activities, and mental states produces substantial errors, whereas tailoring cutoffs to the person and context consistently improves metric validity across datasets and tasks. These results suggest the need for adaptive, signal\bh by\bh signal preprocessing: optimizing filter parameters for each recording can deliver more accurate IBI, RMSSD, and related physiological markers in mental‑ and behavioral\bh health research and real\bh world applications.		



\ifanonymized
 \relax
\else
 \section*{Acknowledgements}
 This research is supported by the \grantsponsor{nih}{National Institutes of Health}{https://doi.org/10.13039/100000002}, under award numbers
 \grantnum{nih}{P30DA029926} and \grantnum{nih}{5R01LM014191-03}, and the \grantsponsor{nsf}{National Science Foundation}{https://doi.org/10.13039/100000001}, under \grantnum{nsf}{IIS-2442593}. ChatGPT was used to correct grammatical errors and improve flow.
\fi 


\bibliographystyle{ACM-Reference-Format}

\bibliography{bibs/local}

\end{document}